\documentclass[runningheads]{llncs}
\usepackage{graphicx}
\usepackage{xcolor}
\usepackage{cite}
\usepackage{subcaption}
\usepackage{wrapfig}
\usepackage{multirow}
\usepackage{pifont}

\usepackage{array}
\newcolumntype{P}[1]{>{\centering\arraybackslash}p{#1}}
\newcolumntype{M}[1]{>{\centering\arraybackslash}m{#1}}

\begin{document}
\title{MRIS: A Multi-modal Retrieval Approach for Image Synthesis on Diverse Modalities}
\titlerunning{MRIS}
%
\author{Boqi Chen, Marc Niethammer}
\authorrunning{B. Chen, M. Niethammer}
%
\institute{Department of Computer Science, University of North Carolina at Chapel Hill
\email{\{bqchen,mn\}@cs.unc.edu}}
\maketitle              
\begin{abstract}
Multiple imaging modalities are often used for disease diagnosis, prediction, or population-based analyses. However, not all modalities might be available due to cost, different study designs, or changes in imaging technology. If the differences between the types of imaging are small, data harmonization approaches can be used; for larger changes, direct image synthesis approaches have been explored. 
In this paper, we develop an approach based on multi-modal metric learning to synthesize images of diverse modalities. We use metric learning via multi-modal image retrieval, resulting in embeddings that can relate images of different modalities. Given a large image database, the learned image embeddings allow us to use k-nearest neighbor ($k$-NN) regression for image synthesis. 
Our driving medical problem is knee osteoarthritis (KOA), but our developed method is general after proper image alignment. We test our approach by synthesizing cartilage thickness maps obtained from 3D magnetic resonance (MR) images using 2D radiographs. Our experiments show that the proposed method outperforms direct image synthesis and that the synthesized thickness maps retain information relevant to downstream tasks such as progression prediction and Kellgren-Lawrence grading (KLG). Our results suggest that retrieval approaches can be used to obtain high-quality and meaningful image synthesis results given large image databases.

\end{abstract}
\section{Introduction}
Recent successes of machine learning algorithms in computer vision and natural language processing suggest that training on large datasets is beneficial for model performance~\cite{radford2021learning,bao2021beit,li2023blip,brown2020language}. While several efforts to collect very large medical image datasets are underway~\cite{littlejohns2019uk,ikram2020objectives}, collecting large \emph{homogeneous} medical image datasets is hampered by: a) cost, b) advancement of technology throughout long study periods, and c) general heterogeneity of acquired images across studies, making it difficult to utilize all data. Developing methods accounting for different imaging types would help make the best use of available data.

Although image harmonization and synthesis~\cite{ren2021segmentation, kasten2020end, kawahara2021t1, boulanger2021deep} methods have been explored to bridge the gap between different types of imaging, these methods are often applied to images of the same geometry. On the contrary, many studies acquire significantly more diverse images; e.g., the OAI image dataset\footnote{https://nda.nih.gov/oai/}~\cite{eckstein2012recent} contains both 3D MR images of different sequences and 2D radiographs. Similarly, the UK Biobank~\cite{littlejohns2019uk} provides different 3D MR image acquisitions and 2D DXA images. Ideally, a machine learning system can make use of all data that is available. As a related first step in this direction, we explore the feasibility of predicting information gleaned from 3D geometry using 2D projection images. Being able to do so would allow a) pooling datasets that drastically differ in image types or b) relating information from a cheaper 2D screening to more readily interpretable 3D quantities that are difficult for a human observer. 

We propose an image synthesis method for diverse modalities based on multi-modal metric learning and $k$-NN regression. To learn the metric, we use image retrieval as the target task, which aims at embedding images such that matching pairs of different modalities are close in the embedding space. We use a triplet loss~\cite{schroff2015facenet} to contrastively optimize the gap between positive and negative pairs based on the cosine distance over the learned deep features. In contrast to the typical learning process, we carefully design the training scheme to avoid interference when training with longitudinal image data. 
Given the learned embedding, we can synthesize images between diverse image types by $k$-NN regression through a weighted average based on their distances measured in the embedding space. Given a large database, this strategy allows for a quick and simple estimation of one image type from another.

We use knee osteoarthritis as the driving medical problem and evaluate our proposed approach using the OAI image data. Specifically, we predict cartilage thickness maps obtained from 3D MR images using 2D radiographs.  
This is a highly challenging task and therefore is a good test case for our approach for the following reasons: 1) cartilage is not explicitly visible on radiographs. Instead, the assessment is commonly based on joint space width (JSW), where decreases in JSW suggest decreases in cartilage thickness~\cite{altman1987radiographic}; 2) the difficulty in predicting information obtained from a 3D image using only the 2D projection data; 3) the large appearance difference between MR images and thickness maps; 4) the need to capture fine-grained details within a small region of the input radiograph. 
While direct regression via deep neural networks is possible, such approaches lack interpretability and we show that they can be less accurate for diverse images.

The main contributions of our work are as follows.
\begin{enumerate}
\item We propose an image synthesis method for diverse modalities based on multi-modal metric learning using image retrieval and $k$-NN regression. We carefully construct the learning scheme to account for longitudinal data.
\item We extensively test our approach for osteoarthritis, where we synthesize cartilage thickness maps derived from 3D MR using 2D radiographs.
\item Experimental results show the superiority of our approach over commonly used image synthesis methods, and the synthesized images retain sufficient information for downstream tasks of KL grading and progression prediction. 
\end{enumerate}

\begin{figure}[t]
  \centering
\includegraphics[width=1.0\textwidth]{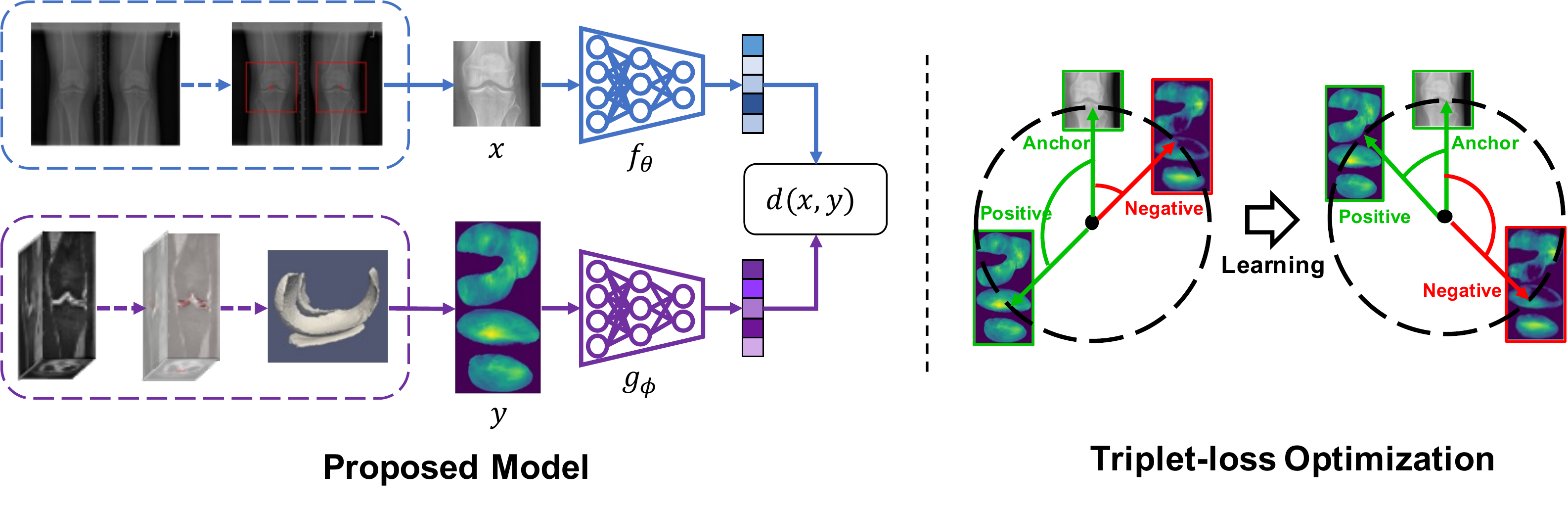}
\caption{Proposed multi-modal metric learning model (left) trained using a triplet loss (right). Left top: encoding the region of interest from radiographs, extracted using the method from~\cite{tiulpin2019kneel}. Left bottom: encoding thickness maps, extracted from MR images using the method from~\cite{huang2022dadp}. Features are compared using cosine similarity. Right: applying triplet loss on cosine similarity, where nonpaired data is moved away from paired data.}
\label{fig:model}
\vspace{-0.3cm}
\end{figure}

\section{Method}
\label{sec:method}
In this work, we use multi-modal metric learning followed by $k$-NN regression to synthesize images of diverse modalities.
Our method requires 1) a database containing matched image pairs; 2) target images aligned to an atlas space. 

\vspace{-0.2cm}
\subsection{Multi-modal Longitudinally-Aware Metric Learning}
\label{subsec:metric_learning}
Let $\{(x_{a}^{i},y_{a}^{i})\}$ be a database of multiple paired images with each pair containing two modalities $x$ and $y$ of the $a$-th subject and $i$-th timepoint if longitudinal data is available. We aim to learn a metric that allows us to reliably identify related image pairs, which in turn relate structures of different modalities.  
Specifically, we train our deep neural network via a triplet loss so that matching image pairs are encouraged to obtain embedding vectors closer to each other than mismatched pairs. Fig.~\ref{fig:model} illustrates the proposed multi-modal metric learning approach, which uses two convolutional neural networks (CNNs), each for extracting the features of one modality. The two networks may share the same architecture, but unlike Siamese networks~\cite{bromley1993signature}, our CNNs have independent sets of weights. This is because the two modalities differ strongly in appearance.

Denoting the two CNNs as $f(\cdot;\theta)$ and $g(\cdot;\phi)$, where $\theta$ and $\phi$ are the CNN parameters, we measure the feature distance between two images $x$ and $y$ using cosine similarity
\begin{equation}
    d(x,y) = 1-\frac{f(x;\theta)\cdot g(y;\phi)}{\left \| f(x;\theta) \right \|\left \| g(y;\phi) \right \|}\,,
\end{equation}
where the output of $f$ and $g$ are vectors of the same dimension\footnote{For notational clarity we will suppress the dependency of $f$ on $\theta$ and will write $f_\theta(\cdot)$ instead of $f(\cdot;\theta)$.}. Given a minibatch of $N$ paired images, our goal is to learn a metric such that $f(x_{a}^{i})$ and $g(y_{a}^{i})$ are close (that is, for the truly matching image pair), while $f(x_a^{i})$ and $g(y_b^{j})$ are further apart, where $a\neq b$ and $i$, $j$ are arbitrary timepoints of subjects $a$, $b$, respectively. \emph{We explicitly avoid comparing across timepoints of the same subject to avoid biasing longitudinal trends.} This is because different patients have different disease progression speeds. For those with little to no progression, images may look very similar across timepoints and should therefore result in similar embeddings. It would be undesirable to view them as negative pairs. Therefore, our multi-modal longitudinally-aware triplet loss becomes
\begin{equation}
    loss( \{(x_a^{i},y_a^{i})\} ) = \sum_{(a,i)} \sum_{(b,j), b\neq a} \max [d(f_\theta(x_a^i),g_\phi(y_a^i))
    -d(f_\theta(x_a^i),g_\phi(y_b^{j})) + m,0]\,,
    \label{eq:longitudinally_aware_triplet_loss}
\end{equation}
where $m$ is the margin for controlling the minimum distance between positive and negative pairs. We sum over all subjects at all timepoints for each batch. 

To avoid explicitly tracking the subjects in a batch, we can simplify the above equation by randomly picking one timepoint per subject during each training epoch. This then simplifies our multi-modal longitudinally aware triplet loss to a standard triplet loss of the form
\begin{equation}
loss( \{(x_a,y_b)\} ) = \sum_{a=1}^N \sum_{b=1,b\neq a}^N \max [d(f_\theta(x_a),g_\phi(y_a))
    -d(f_\theta(x_a),g_\phi(y_b)) + m,0]\,.\nonumber
    \label{eq:standard_triplet}
\end{equation}

\vspace{-0.2cm}
\subsection{Image Synthesis}
\label{subsec:image_synthesis}

\begin{wrapfigure}{r}{0.6\textwidth}
  \centering
  \vspace{-0.8cm}
\includegraphics[width=0.58\textwidth]{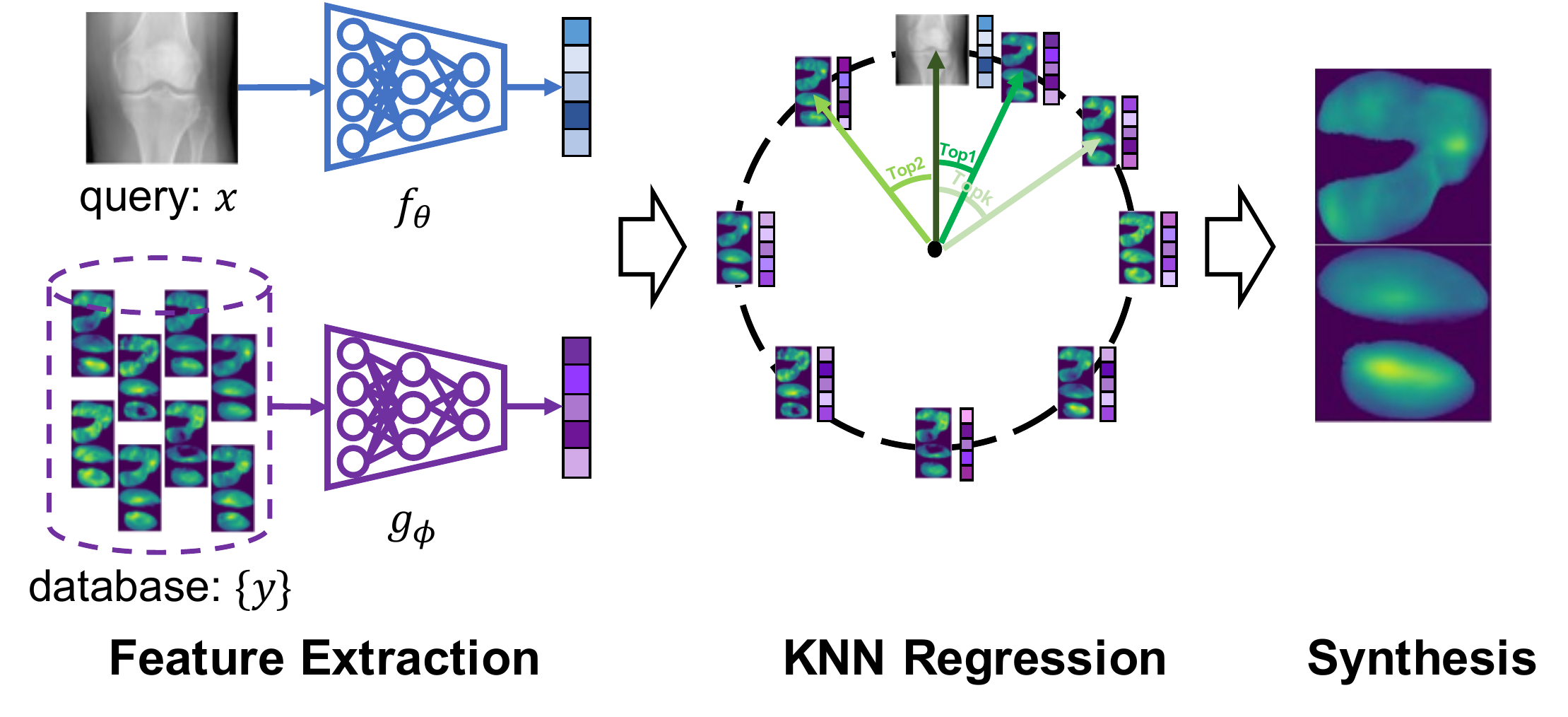}
    \captionof{figure}{Image synthesis by $k$-NN regression from the database. Given an unseen image $x$, we extract its features $f_\theta(x)$, find the $k$ nearest neighbors in the database $\{y\}$ based on these features, and use them for a weighted $k$-NN regression.}
    \label{fig:synthesis}
    \vspace{-0.8cm}
\end{wrapfigure}

After learning the embedding space, it can be used to find the most relevant images with a new input, as shown in Fig.~\ref{fig:synthesis}.
Specifically, the features of a query image $x$ are first extracted by the CNN model $f_\theta$ we described previously. Given a database of images of the target modality $\mathcal{S}^I = \{y_a^{i}\}$ and their respective embeddings $\mathcal{S}^F = \{g(y_a^{i})\}$, we can then select the top $k$ images with the smallest cosine distance, which will be the most similar images given this embedding. Denoting these $k$ most similar images as $\mathcal{K}=\{\tilde{y}^k\}$ we can synthesize an image, $\hat{y}$ based on a query image, $x$ as a weighted average of the form
\begin{equation}
    \hat{y} = \sum_{i=1}^K w_i \tilde{y}^i\quad where \quad w_i = \frac{1-d(x,\tilde{y}^i)}{\sum_{j=1}^K (1-d(x,\tilde{y}^j))}\,,
\end{equation}
where the weights are normalized weights based on the cosine similarities.
This requires us to work in an atlas space for the modality $y$, where all images in the database $\mathcal{S}^I$ are spatially aligned. However, images of the modality $x$ do not need to be spatially aligned, as long as sensible embeddings can be captured by $f_\theta$. 
As we will see, this is particularly convenient for our experimental setup, where the modality $x$ is a 2D radiograph and the modality $y$ is a cartilage thickness map derived from a 3D MR image, which can easily be brought into a common atlas space. As our synthesized image, $\hat{y}$, is a weighted average of multiple spatially aligned images, it will be smoother than a typical image of the target modality. However, we show in Sec.~\ref{sec:experimental_results} that the synthesized images still retain the general disease patterns and retain predictive power.

Note also that our goal is not image retrieval or image reidentification, where one wants to find a known image in a database. 
Instead, we want to synthesize an image for a patient who is not included in our image database. Hence, we expect that no perfectly matched image exists in the database and therefore set $k>1$. Based on theoretical analyses of $k$-NN regression~\cite{chen2018explaining}, we expect the regression results to improve for larger image databases. 

\section{Experimental Results}
\label{sec:experimental_results}

This section focuses on investigating the following questions on the OAI dataset: 
\begin{enumerate}
\item {\it How good is our retrieval performance?} We calculate recall values to determine the performance to retrieve the correct image; 
\item {\it How accurate are our estimated images?} We compare the predicted cartilage thickness maps with those obtained from 3D MR images; 
\end{enumerate}

\begin{enumerate}
\vspace{-0.6cm}
  \setcounter{enumi}{2}
  \item {\it Does our prediction retain disease-relevant information for downstream tasks?} We test the performance of our predicted cartilage thickness maps in predicting KLG and osteoarthritis progressors;
\item {\it How does our approach compare to existing image synthesis models?} We show that our approach based on simple $k$-NN regression compares favorably to direct image synthesis approaches. 
\end{enumerate}

\subsection{Dataset}
\label{subsec:dataset}
We perform a large-scale validation of our method using the Osteoarthritis Initiative (OAI) dataset on almost 40,000 image pairs. This dataset includes $4,796$ patients between the ages of $45$ to $79$ years at the time of recruitment. Each patient is longitudinally followed for up to 96 months. 

\begin{figure}[t]
  \centering
  \includegraphics[width=.9\linewidth]{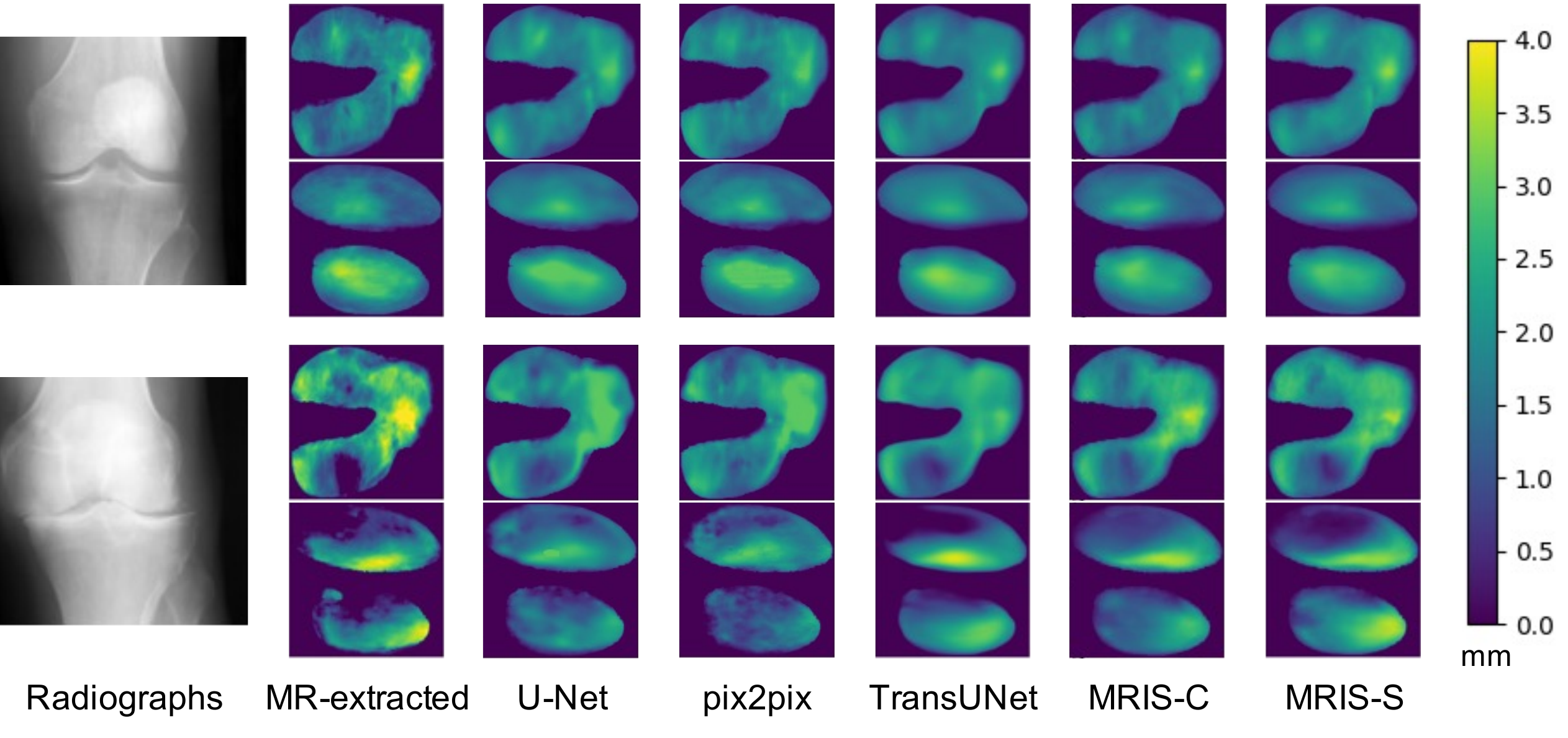}
  \vspace{-0.2cm}
\caption{Thickness map predictions for different methods and different severity. Our approach shows a better match of cartilage thickness with the MR-extracted thickness map than the other approaches. See more examples in the appendix.}
\label{vis_result}
\vspace{-0.5cm}
\end{figure}

\noindent{\bf Images.} The OAI acquired images of multiple modalities, including T2 and DESS MR images, as well as radiographs. We use the paired DESS MR images and radiographs in our experiments. 
After excluding all timepoints when patients do not have complete MR/radiograph pairs, we split the dataset into three sets by patient (i.e., data from the same patient are in the same sets): Set 1) to train the image retrieval model ($2,000$ patients; $13,616$ pairs). This set also acts as a database during image synthesis; Set 2) to train the downstream task ($1,750$ patients; $16,802$ pairs); Set 3) to test performance ($897$ patients; $8,418$ pairs). 
 
\noindent{\bf Preprocessing.} As can be seen from the purple dashed box in Fig.~\ref{fig:model}, we extract cartilage thickness maps from the DESS MR images using a deep segmentation network~\cite{xu2018contextual}, register them to a common 3D atlas space~\cite{shen2019networks}, and then represent them in a common flattened 2D atlas space~\cite{huang2022dadp}. These 2D cartilage thickness maps are our target modality, which we want to predict from the 2D radiographs.
Unlike MR images for which a separate scan is obtained for the left and right knees, OAI radiographs include both knees and large areas of the femur and tibia. To separate them, we apply the method proposed in~\cite{tiulpin2019kneel}, which automatically detects keypoints between the knee joint. As shown in the blue dashed box in Fig.~\ref{fig:model}, the region of interest for each side of the knee is being extracted using a region of $140~mm*140~mm$ around the keypoints.

\begin{wraptable}{r}{0.57\textwidth}
\vspace{-0.5cm}
\resizebox{0.57\textwidth}{!}{
\centering
    \begin{tabular}{|c||P{1.55cm}|P{1.55cm}|P{1.55cm}|P{1.55cm}|}
         \hline
          Method & R@1~$\uparrow$ & R@5~$\uparrow$ & R@10~$\uparrow$ & R@20~$\uparrow$  \\
         \hline
          Femoral & $28.26$ & $58.19$ & $71.13$ & $82.11$   \\
         \hline
          Tibial & $30.49$ & $61.48$ & $73.36$ & $83.33$  \\
         \hline
          Combined & $\textbf{45.21}$ & $\textbf{75.53}$ & $\textbf{84.73}$ & $\textbf{90.64}$ \\
         \hline
    \end{tabular}}
    \caption{Thickness map retrieval recall percentage on the testing set. R@k shows the percentage of queries for which the correct one is retrieved within the top $k$ nearest neighbors.}
    \label{recall}
\vspace{-0.5cm}
\end{wraptable}

We normalize all input radiographs by linearly scaling the intensities so that the smallest $99\%$ values are mapped to $\left[0, 0.99\right]$. We horizontally flip all right knees to the left as done in~\cite{huang2022dadp}, randomly rotate images up to 15 degrees, add Gaussian noise, and adjust contrast. Unlike the radiographs, we normalize the cartilage thickness map by dividing all values by 3, which is approximately the 95-th percentile of cartilage thickness. All images are resized to $256*256$.

\subsection{Network training}
\label{subsec:training}
During multi-modal metric learning, our two branches use the ResNet-18~\cite{he2016deep} model with initial parameters obtained by ImageNet pre-training~\cite{deng2009imagenet}. We fine-tune the networks using AdamW~\cite{loshchilov2017decoupled} with initial learning rate $10^{-4}$ for radiographs and $10^{-5}$ for the thickness maps. 
The output embedding dimensions of both networks are $512$.
We train the networks with a batch size of $64$ for a total of $450$ epochs with a learning rate decay of $80\%$ for every $150$ epochs. We set the margin $m=0.1$ in all our experiments. 

For both downstream tasks, we fine-tune our model on a ResNet-18 pre-trained network with the number of classes set to $4$ for KLG prediction and $2$ for progression prediction. Both tasks are trained with AdamW for $30$ epochs, batch size $64$, and learning rate decay by $80\%$ for every $10$ epochs. The initial learning rate is set to $10^{-5}$ for KLG prediction and $10^{-4}$ for progression prediction.

\subsection{Results}
\label{subsec:results}
This section shows our results for image retrieval, synthesis, and downstream tasks based on the questions posed above. All images synthesized from MRIS are based on the weighted average of the retrieved top $k=20$ thickness maps.

\begin{table}[t]
    \begin{center}
    \resizebox{\textwidth}{!}{
    \begin{tabular}{|M{1.3cm}|M{1.6cm}||M{2.0cm}|M{2.0cm}|M{2.0cm}|M{2.0cm}||M{2.0cm}|}
        \hline
        \multicolumn{2}{|c||}{Median $\pm$ MAD} & KLG01~$\downarrow$ & KLG2~$\downarrow$ & KLG3~$\downarrow$ & KLG4~$\downarrow$ & All~$\downarrow$ \\
        \hline
        \multirow{5}{*}{\shortstack{Femoral\\ Cartilage}} & U-Net & $0.288 \pm 0.173$ & $0.324 \pm 0.195$ & $0.358 \pm 0.214$ & $0.410 \pm 0.252$ & $0.304 \pm 0.183$\\
        \cline{2-7}
        & pix2pix & $0.289 \pm 0.173$ & $0.326 \pm 0.196$ & $0.360 \pm 0.216$ & $0.411 \pm 0.253$ & $0.306 \pm 0.183$ \\
        \cline{2-7}
        & TransUNet & $0.260 \pm 0.157$ & $0.300 \pm 0.180$ & $0.326 \pm 0.195$ & $0.384 \pm 0.235$ & $0.277 \pm 0.167$ \\
        \cline{2-7}
        & MRIS-C & $0.265 \pm 0.158$ & $0.298 \pm 0.178$ & $0.319 \pm 0.191$ & $0.377 \pm 0.226$ & $0.279 \pm 0.167$ \\
        \cline{2-7}
        & MRIS-S & $\textbf{0.259} \pm \textbf{0.155}$ & $\textbf{0.295} \pm \textbf{0.176}$ & $\textbf{0.319} \pm \textbf{0.191}$ & $\textbf{0.373} \pm \textbf{0.223}$ & $\textbf{0.275} \pm \textbf{0.164}$ \\
        \hline
        \hline
        \multirow{5}{*}{\shortstack{Tibial\\ Cartilage}} & U-Net & $0.304 \pm 0.181$ & $0.324 \pm 0.193$ & $0.364 \pm 0.216$ & $0.428 \pm 0.270$ & $0.316 \pm 0.188$ \\
        \cline{2-7}
        & pix2pix & $0.306 \pm 0.182$ & $0.325 \pm 0.194$ & $0.367 \pm 0.219$ & $0.433 \pm 0.272$ & $0.319 \pm 0.190$ \\
        \cline{2-7}
        & TransUNet & $0.269 \pm 0.160 $ & $0.288 \pm 0.172$ & $0.325 \pm 0.192$ & $\textbf{0.371} \pm \textbf{0.254}$ & $0.281 \pm 0.167$ \\
        \cline{2-7}
        & MRIS-C & $0.271 \pm 0.160$ & $0.291 \pm 0.171$ & $0.319 \pm 0.188$ & $0.385 \pm 0.225$ & $0.282 \pm 0.166$ \\
        \cline{2-7}
        & MRIS-S & $\textbf{0.265} \pm \textbf{0.157}$ & $\textbf{0.283} \pm \textbf{0.168}$ & $\textbf{0.313} \pm \textbf{0.187}$ & $0.379 \pm 0.226$ & $\textbf{0.276} \pm \textbf{0.163}$ \\
        \hline
    \end{tabular}
    }
    \end{center}
    \caption{Median $\pm$ MAD absolute error for both femoral and tibial cartilage between the predicted thickness maps and those extracted from MR images. We stratify the result by KLG. Larger KLG results in less accurate synthesis.}
    \label{abs_error}
    \vspace{-0.7cm}
\end{table}

\noindent{\bf Image retrieval.} To show the importance of the learned embedding space, we perform image retrieval on the test set, where our goal is to correctly find the corresponding matching pair. Since our training process does not compare images of the same patient at different timepoints, we test using only the baseline images for each patient ($1,794$ pairs). During training, we created two thickness map variants: 1) combining the femoral and tibial cartilage thickness maps (Combined); 2) separating the femoral and tibial thickness maps (Femoral/Tibial), which requires training two networks. 
Tab.~\ref{recall} shows the image retrieval recall, where R@k represents the percentage of radiographs for which the correct thickness map is retrieved within the $k$-nearest neighbors in the embedding space. Combined achieves better results than retrieving femoral and tibial cartilage separately. This may be because more discriminative features can be extracted when both cartilages are provided, which simplifies the retrieval task. In addition, tibial cartilage appears to be easier to retrieve than femoral cartilage.

\noindent{\bf Image synthesis.} To directly measure the performance of our synthesized images on the testing dataset, we show the median $\pm$ MAD (median absolute deviation) absolute error compared to the thickness map extracted by MR in Tab.~\ref{abs_error}. We created two variants by combining or separating the femoral and tibial cartilage, corresponding to MRIS-C(ombined) and MRIS-S(eparate). Unlike the image retrieval recall results, MRIS-S performs better than MRIS-C (last column of Tab.~\ref{abs_error}). This is likely because it should be beneficial to mix and match separate predictions for synthesizing femoral and tibial cartilage. Moreover, MRIS-S outperforms all baseline image synthesis methods~\cite{U-Net, pix2pix, chen2021transunet}.  

Osteoarthritis is commonly assessed via Kellgren-Lawrence grade~\cite{kellgren1957radiological} on radiographs by assessing joint space width and the presence of osteophytes. KLG=$0$ represents a healthy knee, while KLG=$4$ represents severe osteoarthritis. KLG=$0$ and $1$ are often combined because knee OA is considered definitive only when KLG$\geq2$~\cite{kohn2016classifications}. To assess prediction errors by OA severity, we stratify our results in Tab.~\ref{abs_error} by KLG. Both variants of our approach perform well, outperforming the simpler pix2pix and U-Net baselines for all KLG. The TransUNet approach shows competitive performance, but overall our MRIS-S achieves better results regardless of our much smaller model size. Fig.~\ref{vis_result} shows examples of images synthesized for the different methods for different severity of OA.

\begin{table}[t]
    \begin{center}
    \scalebox{0.95}{
    \begin{tabular}{|c||M{1.2cm}|M{1.2cm}|M{1.2cm}|M{1.2cm}|M{1.2cm}||M{1.5cm}|M{1.5cm}|}
         \hline
          \multirow{2}{*}{Method} & \multicolumn{5}{c||}{KLG Prediction (accuracy)~$\uparrow$} & \multicolumn{2}{c|}{Progression Prediction} \\
         \cline{2-8}
          & KLG01 & KLG2 & KLG3 & KLG4 & overall & average precision~$\uparrow$ & roc auc~$\uparrow$ \\
          \hline
          U-Net & $0.819$ & $0.321$ & $0.778$ & $0.545$ & $0.719$  & $0.242$ & $0.606$\\
         \hline
          pix2pix  & $0.805$ & $0.396$ & $0.735$ & $0.654$ & $0.722$ & $0.225$ & $0.625$\\
         \hline
         TransUNet & $0.797$ & $\textbf{0.528}$ & $\underline{0.763}$ & $\textbf{0.865}$ & $0.746$ & $0.286$ & $0.654$ \\
         \hline
         MRIS-C  & $\underline{0.865}$ & $0.469$ & $0.757$ & $0.673$ & $\underline{0.781}$ & $\underline{0.299}$ & $\underline{0.713}$\\
         \hline
          MRIS-S  & $\textbf{0.869}$ & $0.479$ & $\textbf{0.786}$ & $0.718$ & $\textbf{0.789}$ & $\textbf{0.307}$  & $0.702$\\
         \hline
         MR-extracted & $0.842$ & $\underline{0.523}$ & $0.727$ & $\underline{0.795}$ & $0.775$ & $0.286$ & $\textbf{0.739}$ \\
         \hline
    \end{tabular}
    }
    \end{center}
    \caption{Results on the downstream tasks of KLG and progression prediction. Our synthesis methods overall perform better than other synthesis methods and obtain a comparable result with the MR-extracted thickness maps.}
    \label{downstream_task}
    \vspace{-0.6cm}
\end{table}

\noindent{\bf Downstream tasks.} The ultimate question is whether the synthesized images can still retain information for downstream tasks. Therefore, we test the ability to predict KLG 
and OA progression, where we define OA progression as whether or not the KLG will increase within the next 72 months.
Tab.~\ref{downstream_task} shows that our synthesized thickness maps perform on par with the MR-extracted thickness maps for progression prediction and we even outperform on predicting KLG. MRIS overall performs better than U-Net~\cite{U-Net}, pix2pix~\cite{pix2pix} and TransUNet~\cite{chen2021transunet}.

\section{Conclusion}
\label{sec:conclusion}
In this work, we proposed an image synthesis method using metric learning via multi-modal image retrieval and $k$-NN regression. We extensively validated our approach using the large OAI dataset and compared it with direct synthesis approaches. We showed that our method, while conceptually simple, can effectively synthesize alignable images of diverse modalities. More importantly, our results on the downstream tasks showed that our approach retains disease-relevant information and outperforms approaches based on direct image regression. 
Potential shortcomings of our approach are that the synthesized images tend to be smoothed due to the weight averaging and that spatially aligned images are required for the modality to be synthesized. 

\section{Acknowledgements}
This work was supported by NIH 1R01AR072013; it expresses the views of the authors, not of NIH. Data and research tools used in this manuscript were obtained / analyzed from the controlled access datasets distributed from the Osteoarthritis Initiative (OAI), a data repository housed within the NIMH Data Archive. OAI is a collaborative informatics system created by NIMH and NIAMS to provide a worldwide resource for biomarker identification, scientific investigation and OA drug development. Dataset identifier: NIMH Data Archive Collection ID: 2343.

%
%
%
\bibliographystyle{splncs04}
\bibliography{main}

\clearpage
\appendix


\section{Cartilage Thickness Longitudinal Trend}
\vspace{-0.4cm}
\begin{figure}
\centering
\includegraphics[width=.7\linewidth]{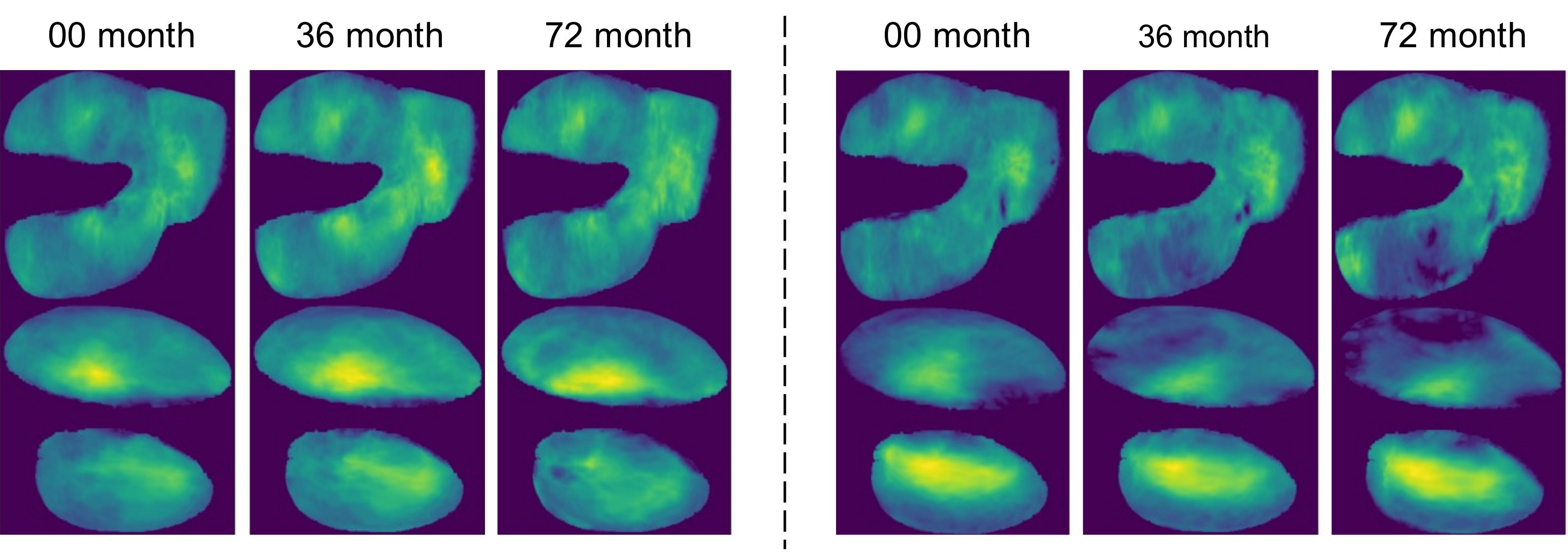}
\caption{Example thickness maps of two patients over a period of 72 months. The left patient is a non-progression patient with KLG=$0$ for all timepoints, while the right patient is a progressor with KLG=$1$ at baseline and KLG=$3$ at 72 months. Changes for the non-progression patient are subtle. Hence, different timepoints of one patient should not be used as negative examples for the triplet loss.}
\end{figure}

\vspace{-0.4cm}
\section{Examples for Different Methods and KLGs}
\vspace{-0.4cm}
\begin{figure}[h!]
  \centering
  \includegraphics[width=.95\linewidth]{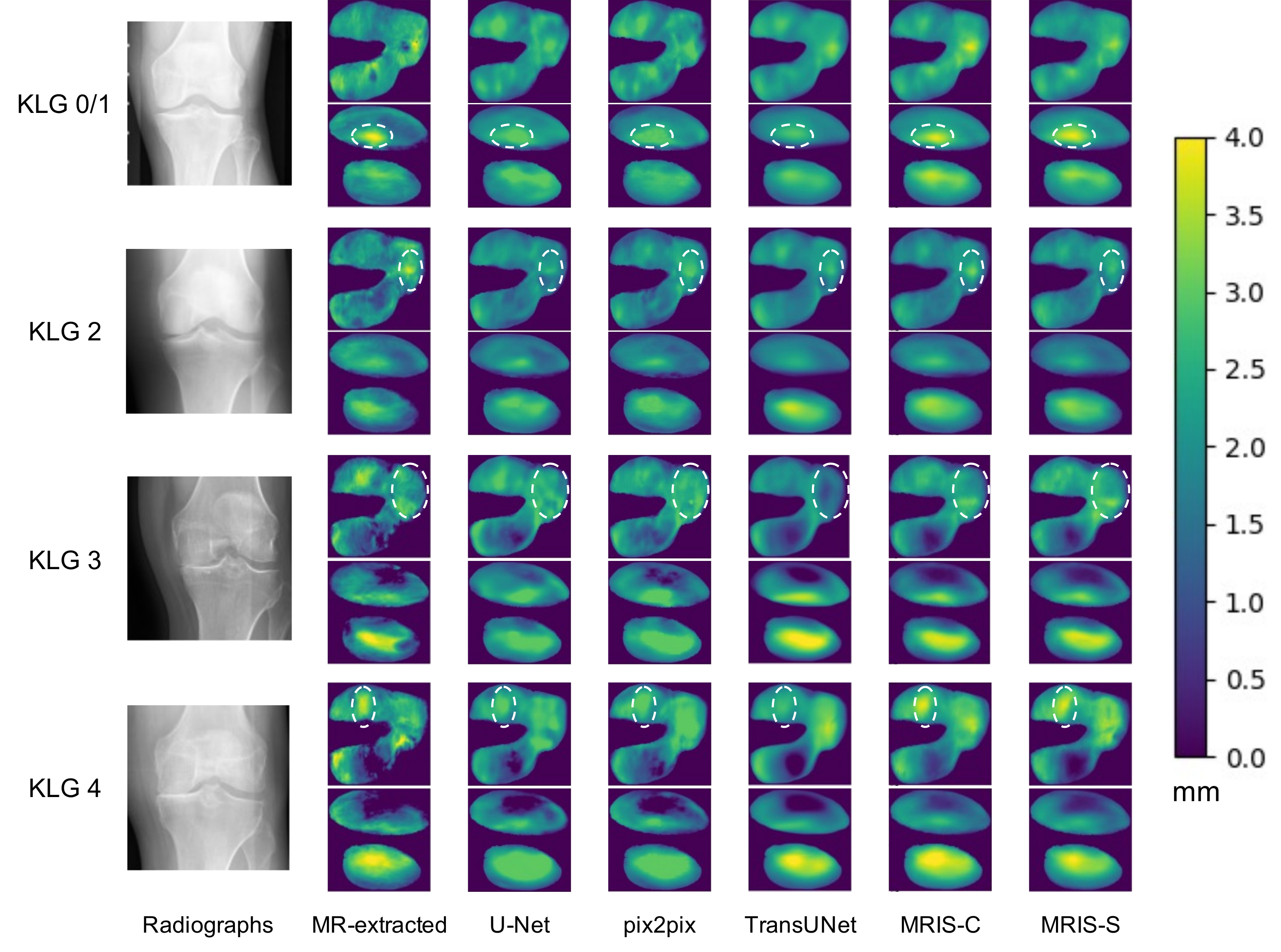}
\caption{Predicted thickness map for different methods and KLGs. As KLG increases, the predicted thickness map becomes less accurate, but our method can generate patterns that better match the MR-extracted thickness map.}
\label{result_more}
\end{figure}

\clearpage
\section{Failure Cases}
\vspace{-0.4cm}
\begin{figure}[h!]
  \centering
  \includegraphics[width=.99\linewidth]{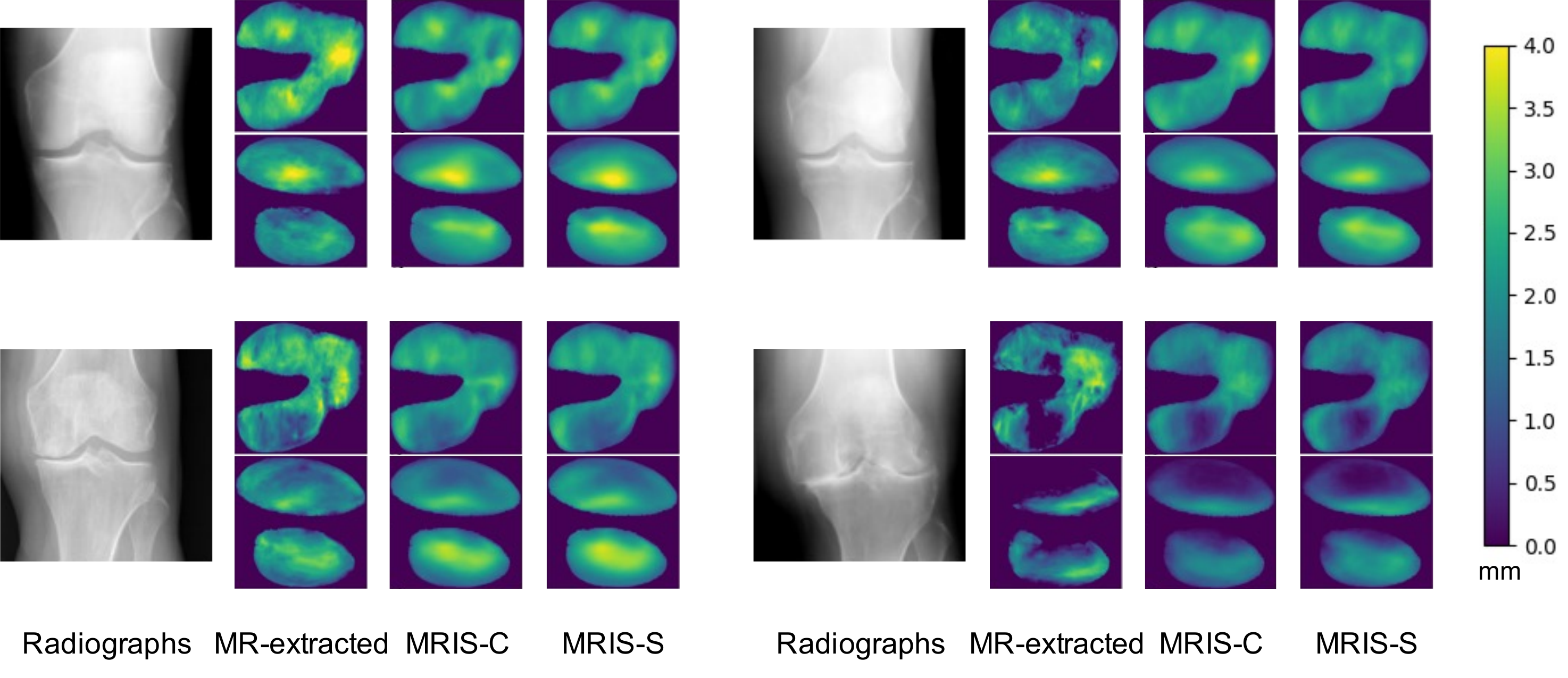}
\caption{Examples of predicted thickness maps that failed to capture the thickness pattern. The examples to the left mainly failed to capture the tibial cartilage pattern (bottom), and the examples to the right mainly failed to capture the femoral cartilage pattern (top). Overall, our MRIS model is less accurate for generating thickness maps with severe diseases. This may be due to the limited number of samples with such severe cartilage patterns in our database.}
\label{result_more}
\end{figure}

\vspace{-0.4cm}
\section{Model Size Comparison}
\vspace{-0.4cm}
\begin{table}[h!]
    \begin{center}
    \scalebox{0.95}{
    \begin{tabular}{|M{2.5cm}||M{2.5cm}|M{2.5cm}|}
    \hline
    Model & \# params & flops \\
    \hline
    U-Net & $54.403$ M & $17.844$ G \\
    \hline
    pix2pix & $57.167$ M & $20.977$ G \\
    \hline
    TransUNet & $93.231$ M & $24.670$ G\\
    \hline
    MRIS & $\textbf{22.353}$ \textbf{M} & $\textbf{4.763}$ \textbf{G} \\
    \hline
    \end{tabular}
}
    \end{center}
    \caption{Model size comparison between different methods. We calculate the number of parameters and flops for each model. Even though our MRIS model is the smallest, we can still achieve a better synthesis result that can be meaningful for downstream tasks.}
    \label{model_compare}
\end{table}
\end{document}